\documentclass[twocolumn]{aastex62}
\submitjournal{ApJ}
\usepackage{enumitem}
\usepackage{graphicx,url}
\usepackage{color}
\usepackage{amsmath}
\def\gx339{GX~339$-$4}

\def\reflionx{{\tt reflionx}}
\def\xillver{{\tt xillver}}

\def\deg{$^{\circ}$}

\shorttitle{Accurate Comptonization in X-ray Reflection Models}
\shortauthors{Garc\'{\i}a \& et al.}

\begin{document}

%\title{\large\bf Accurate Treatment of Comptonization in X-ray Reflection Models}
\title{\large\bf Accurate Treatment of Comptonization in X-ray Illuminated Accretion Disks}

\correspondingauthor{Javier~A.~Garc\'ia}
\email{javier@caltech.edu}

\author[0000-0003-3828-2448]{Javier~A.~Garc\'ia}
\affil{Cahill Center for Astronomy and Astrophysics, California Institute of Technology, Pasadena, CA 91125, USA}
\affil{Dr. Karl Remeis-Observatory and Erlangen Centre for Astroparticle Physics, Sternwartstr.~7, 96049 Bamberg, Germany}

\author{Ekaterina~Sokolova-Lapa}
\affil{Dr. Karl Remeis-Observatory and Erlangen Centre for Astroparticle Physics, Sternwartstr.~7, 96049 Bamberg, Germany}
\affil{Sternberg Astronomical Institute, M.~V.~Lomonosov Moscow State University,Universitetskij pr., 13, Moscow 119992, Russia}

\author{Thomas~Dauser}
\affil{Dr. Karl Remeis-Observatory and Erlangen Centre for Astroparticle Physics, Sternwartstr.~7, 96049 Bamberg, Germany}

\author{Jerzy~Madej}
\affil{Astronomical Observatory, University of Warsaw, Al. Ujazdowskie 4, 00-478 Warszawa, Poland}

\author{Agata~R\'o\.za\'nska}
\affil{Nicolaus Copernicus Astronomical Center, Polish Academy of Sciences, Bartycka 18, 00-716 Warszawa, Poland}

\author{Agnieszka~Majczyna}
\affil{National Centre for Nuclear Research, ul. Andrzeja So\l{}tana 7, 05-400 Otwock, Poland}

\author{Fiona~A.~Harrison}
\affil{Cahill Center for Astronomy and Astrophysics, California Institute of Technology, Pasadena, CA 91125, USA}

\author{J\"orn~Wilms}
\affil{Dr. Karl Remeis-Observatory and Erlangen Centre for Astroparticle Physics, Sternwartstr.~7, 96049 Bamberg, Germany}

%==================================================================================
\begin{abstract}
A large fraction of accreting black hole and neutron stars systems present
clear evidence of the reprocessing of X-rays in the atmosphere of an
optically-thick accretion disk.  The main hallmarks of X-ray reflection include
fluorescent K-shell emission lines from iron ($\sim 6.4-6.9$\,keV), the absorption
iron K-edge ($\sim 7-9$\,keV), and a broad featureless component known as the
{\it Compton hump} ($\sim 20-40$\,keV). This Compton hump is produced as the
result of the scattering of high-energy photons ($E \gtrsim 10$\,keV) of the
relatively colder electrons ($T_e \sim 10^5-10^7$\,K) in the accretion disk, in
combination with photoelectric absorption from iron. The treatment of this
process in most current models of ionized X-ray reflection has been done using
an approximated Gaussian redistribution kernel. This approach works
sufficiently well up to $\sim100$\,keV, but it becomes largely inaccurate at
higher energies and at relativistic temperatures ($T_e\sim10^9$\,K). We present
new calculations of X-ray reflection using a modified version of our code
\xillver, including an accurate solution for Compton scattering of the
reflected unpolarized photons in the disk atmosphere. This solution takes into account
quantum electrodynamic and relativistic effects allowing the correct treatment
of high photon energies and electron temperatures. We show new reflection
spectra computed with this model, and discuss the improvements achieved in the
reproducing the correct shape of the Compton hump, the discrepancies with
previous calculations, and the expected impact of these new models in the
interpretation of observational data.
\end{abstract}


%
%==================================================================================
%
%
%
\section{Introduction}\label{sec:intro}

Accretion onto compact objects such as black holes or neutron stars is one of
the most efficient mechanisms to convert gravitational energy into radiation.
This radiation is mostly comprised of very energetic photons, making X-ray
spectroscopy a resourceful technique to study these systems, and their
interaction with the surrounding material. In the case of black holes, the
X-ray continuum is typically dominated by a non-thermal emission in the form of
a power-law that extends to high energies, which is thought to be produced
either in a centrally located and  hot ($T_e\sim 10^9$\,K) plasma, which origin
is still a matter of debate \citep[e.g.,][]{sha73,haa93,mat92,mar05}.  Thermal
emission from the accretion disk can peak from the ultraviolet band to the soft
X-rays, depending on the mass of the black hole.

A fraction of the non-thermal emission illuminates the accretion disk,
producing a rich reflection spectrum of fluorescent lines and other spectral
features. This reflection component can appear ionized, with the most prominent
spectral lines being due to Fe K emission at 6.4--6.9\,keV
\citep[e.g.,][]{ros05,gar10}. These features can also be severely distorted in
the strong-gravity regime by the Doppler effect, light bending, and
gravitational redshift \citep[e.g.,][]{fab89,lao91}.  Relativistically
broadened Fe K lines have been observed in the spectra of the majority of
well-studied black hole binaries, as well as in a large fraction of active galactic nuclei
\cite[e.g.,][]{bre13b,rey19}.

At energies well above the Fe K-threshold ($\sim7-9$\,keV), the photoelectric
cross section of the metals decreases rapidly and electron scattering becomes
the dominant source of opacity. Thus, high-energy photons will suffer several
scatterings with the electrons in the upper layers of the accretion disk.
Here, we only consider high-energy photons with energies less than 1.022\,MeV
(twice the electron rest mass), because we do not treat electron-positron pair
production.  These photons lose energy after every scattering, roughly a Compton
wavelength per event ($\lambda_\mathrm{C}=h/m_ec \approx 0.024$\AA, where $h$
is the Planck's constant, $m_e$ the electron rest mass, and $c$ the speed of
light).  Since electrons are not at rest (particularly if the gas temperature
$T_e$ is high), photons will lose an additional fractional energy per scattering
($\Delta E/E \sim 3kT_e/m_ec^2$).  The reduction of the number of photons at
high energies due to electron scattering, and at lower energies due to the
photoelectric opacity of iron, leads to a broad and featureless spectral
feature centered around 20--30\,keV, typically referred to as the {\it Compton
hump} \citep{lig88,gui88}. This distinct feature is considered one of the hallmarks of X-ray
reflection in optically-thick media. 

A detailed calculation of the redistribution of high-energy photons due
to Compton scattering is crucial to correctly predict the detailed spectral
shape of X-ray reprocessed in accretion disks. One of the most accurate methodologies
to solve the Comptonization problem is via Monte Carlo calculations, where individual
photons are ``followed" as they interact with the electrons in the gas \citep[e.g.][]{geo91,mat91,mat93}.
The Monte Carlo method has the advantage that it treats the radiative transfer problem
exactly (within the numerical precision), given that no other source of
opacity is consider. However, the drawback is that computationally expensive, and
becomes prohibitive when the complexity of the microphysics is increased (e.g.,
when atomic lines are also considered). Thus, in problems where the coupling between
the radiation and the medium is described with enough physics, the price to pay
is the implementation of more approximated methods to solve the radiation transfer.

The complete problem of reflected (reprocessed) X-rays in optically-thick media 
such as accretion disks has been subject of study for nearly four decades now. Among
the most popular models are \reflionx\ \citep{ros05} and \xillver\ \citep{gar10,gar11,gar13a}. 
Until now, these models have treated the redistribution of photons due to Compton
scattering in a rather approximated fashion, using a simple redistribution function
based on a Gaussian profile \citep[e.g.][]{ros93,nay00,gar10}. This approximation allows
for a fast calculation of the photon redistribution at each depth in the atmosphere,
which is particularly challenging when the required energy resolution of the model
is high. However, the Gaussian approximation becomes increasingly inaccurate at
energies close or above the electron rest mass energy ($m_ec^2=511$\,keV), or at
very high temperatures where the thermal motion of electrons becomes relativistic,
which causes a suppression of the scattering cross section below the Klein--Nishina
prediction \citep{mad17}.

The redistribution function accurate for photon energies approaching electron
rest mass, was given by \citet{guilbert1981} and used by \citet{madej2000} and
\citet{madej2004} in case of irradiated stellar atmospheres in hydrostatic and
radiative equilibrium. Furthermore, this procedure was adopted to the
stratified accretion disk atmosphere, but with limited number of atomic
opacities \citep{rozanska2008,rozanska2011}.

In this paper we present new calculations of the redistribution function due to electron
scattering as a function of photon energy and electron temperature, implementing an
accurate solution that takes into account quantum electrodynamic and relativistic effects,
allowing the correct treatment of high photon energies and electron temperatures. A
detailed comparison with the standard Gaussian approximation is also presented. Moreover,
we implement this new solution into our reflection model \xillver\ and discuss the
discrepancies with previous calculations, and the expected impact of these new models
in the interpretation of observational data.

The remainder of the paper is organized as follows. In Section~\ref{sec:numeric} we outline the basic approach
for the radiative transfer calculations, and describe in detail the expressions used for the redistribution
of photons due to Compton scattering. In Section~\ref{sec:results} we show the main results, including a comparison
of the Gaussian and exact solutions in the convolution of a single radiation field, several iterations,
and reflection calculations done at different levels of complexity. A discussion of these calculations
and our main conclusions are presented in Section~\ref{sec:disc}.

%----------------------------------------------------------------------------------
\vspace{2cm}
\section{Numerical Calculations}\label{sec:numeric}

\subsection{Radiative Transfer}

There exist extensive literature on the problem of radiation transfer in opaque
media. The reader is referred to review such works, in particular those by
\cite{cha60}, \cite{mih78}, and \cite{hub14}, which are considered the seminal works
in the field.  Here, we only discuss the general equations in order to describe
the problem in hand. 

In the one dimensional case applicable to plane-parallel atmospheres, the
standard form of the radiative transfer equation in steady state can be written
as:
\begin{equation}\label{eq:rt}
\mu \frac{\partial I(E,\mu)}{\partial\tau(E)} = I(E,\mu) - S(E)
\end{equation}
with $\mu=\cos\theta$, where $\theta$ is the angle between the direction of propagation
of the intensity $I(E,\mu)$ and the spatial coordinate. In Equation~\ref{eq:rt} we have
omitted the explicit dependence on the optical depth $\tau$, which is the energy dependent
opacity $\chi(E)$ of the material along the line of sight:
\begin{equation}
\tau(E) = \int{- \chi(E) dz}.
\end{equation}
In general, both absorption and scattering processes contribute to the total opacity,
$\chi(E) = n_{\rm e}(\sigma_a + \sigma_s)$, where $n_e$ is the electron number density,
$\sigma_a$ and $\sigma_s$ are the absorption and scattering cross sections, respectively.

The source function $S(E)$ is defined as the ratio of the total emissivity,
$\eta(E)=n_{\rm e}[\sigma_a B(E,T) + \sigma_s J_c(E)]$, to the total opacity:
\begin{equation}
S(E) = \frac{\sigma_a B(E,T) + \sigma_s J_c(E)}{\sigma_a + \sigma_s}
\end{equation}
Here, the first term represents the gas emissivity assuming local thermal equilibrium,
where $B(E,T)$ is the Planck's function at the local temperature. The
second term is the emissivity due to electron scattering\footnote{In general, other
scattering events could be considered, such as Rayleigh scattering due to molecules,
but in the context of X-ray illuminated accretion disks these are negligible.}, which
is proportional to the Comptonized mean intensity of the radiation field, 
resulting from the convolution
\begin{equation}
J_c(E) = \frac{1}{\sigma_s}\int{dE'J(E')R(E',E)},
\end{equation}
where $J(E)=\frac{1}{2}\int_{-1}^{+1}{d\mu I(E,\mu)}$ is the first moment of the 
radiation field, and $R(E',E)$ is the {\it redistribution function}, which represents
the probability that a photon will be scattered from an energy $E'$ to an energy
$E$. For the proposes of this paper, we consider this redistribution to be isotropic,
i.e., independent of the direction of the incoming and outgoing photon. The presence
of scattering terms in the radiative transfer equation is one of the main difficulties
in the solution of real physical problems, as it decouples the radiation field from
the local properties of the material, allowing photons to be transported over large
distances.

\subsection{The Redistribution Function}

After an inelastic scattering event with an electron, an X-ray photon can gain or 
lose energy. The energy exchange depends on the initial photon energy, and on the
electron temperature. The probability for a photon with energy $E_i$ to have an
energy $E_f$ after the scattering is given by the {\it redistribution function},
$R(E_i, E_f)$. Integration of this function over the final
photon energies results in the energy dependent Compton scattering cross section:
\begin{equation}
\sigma_{\mathrm{CS}}(E_{i}) = \sigma_{\mathrm{T}}\int R(E_{i}, E_{f})dE_{f},
\end{equation}
where $\sigma_{\mathrm{T}}\approx6.65\times10^{-25}\,\mathrm{cm}^2$ is the classical 
Thomson scattering cross section \citep{thom1906}.

There is a number of different redistribution functions used in the literature
to describe the probability of photon with initial energy $E_i$ to be scattered
off the electron with the final energy $E_f$. The most complete and detailed
review on the different approximations for Compton scattering redistribution
functions and their limitations is given by \cite{mad17}. Here, we will focus
on two of the most relevant forms: the exact quantum mechanical formula for
relativistic electrons; and the Gaussian approximation for thermal Compton
scattering, as first implemented in the problem of X-ray reflection by
\cite{ros78}.  Next, these two solutions are described in detail.

\subsubsection{The Exact Redistribution Function}

The first expression for the exact formula of the redistribution function was
obtained by \cite{jon1968}, and later corrected by \citet{ahar1981}. Since
then, a number of works have discussed different ways to simplify the given
equation and to perform an accurate integration
\citep[e.g.,][]{ker1986,nag93,napo1994}.  Here, we adopt the second exact form
given by \cite{nag93} (hereafter NP93),
\begin{equation}\label{redexg}
\begin{aligned}
R_\mathrm{E}(x, x_1, \mu, \gamma) &= 
\frac{2}{Q} + \frac{u}{v}\left( 1-\frac{2}{q}\right) + \\
& u\frac{(u^2-Q^2)(u^2+5v)}{2q^2v^3} + u\frac{Q^2}{q^2v^2},
\end{aligned}
\end{equation}
where $x$, $x_1$ are the dimensional photon energies before and after the scattering, 
$\mu$ is the cosine of the scattering angle, $\gamma$ is the electron Lorenz factor, 
$q=xx_1(1-\mu)$, and $Q^2=(x-x_1)^2+2q$. The functions $u$ and $v$ are defined through 
\begin{equation}
a^2_{-}=(\gamma - x)^2 + \frac{1+\mu}{1-\mu}
\end{equation}
and
\begin{equation}
a^2_{+}=(\gamma + x_1)^2 + \frac{1+\mu}{1-\mu},
\end{equation}
as following,
\begin{equation}
u = a_{+}-a_{-}=\frac{(x+x_1)(2\gamma+x_1-x)}{a_{-}+a_{+}}
\end{equation}
and
\begin{equation}
v=a_{-}a_{+},
\end{equation}
in order to avoid the accuracy loss due to numerical cancellation \citep[see the detailed overview given 
by][Sec.~3.2]{mad17}. The resulting redistribution function for Compton scattering is 
obtained by integrating Eq.~\ref{redexg} with the relativistic Maxwellian distribution
\begin{equation}\label{redex}
\begin{aligned}
R_\mathrm{E}(x, x_1, \mu) = & \frac{3}{32\mu\Theta K_{2}(1/\Theta)} \\ 
&   \int_{\gamma_{*}}^{\infty}R_\mathrm{E}(x, x_1, \mu, \gamma)\exp(-\gamma/\Theta)d\gamma, \\
\end{aligned}
\end{equation}
where $K_{2}$ is a modified Bessel function of the second kind (the Macdonald's
function), and the lower limit of the integral is $\gamma_{*}=(x - x_1 +
Q\sqrt{1+2/q)}/2$ \footnote{We note that there is a misprint in the definition of
$\gamma_{*}$ in \cite{mad17}, Eq.~4. See Erratum by \cite{mad19}.}. We refer to
the redistribution function given by Eq.~\ref{redex} as {\it exact} in the
following, in the sense that it includes all the necessary physical effects,
though still it is not free of the computational errors inserted by numerical
calculations. We also remind the fact that we neglect pair production for
energies greater than 1.022\,MeV.

\subsubsection{The Gaussian Redistribution Function}

In the case of relatively low photon energies ($E \ll m_ec^2$), and for
electron temperatures much lower than the photon field ($kT_e \ll E$), a
simple approximation can be implemented, in which scattered photons are assumed
to be distributed according to a Gaussian profile. This idea was first
introduced by \cite{dir25} and later by \cite{mun48} in order to take into
consideration the velocity of thermal agitation of electrons in stellar
atmospheres. 

As described in \cite{ros93}, the probability for a photon with initial energy $E_i$
to be scattered to a final energy $E_f$ can be written as
\begin{equation}
P(E_{i}, E_{f}) = \frac{1}{\sqrt{2\pi}\Sigma}\exp\bigg[-\frac{-(E_{f}-E_{c})^{2}}{2\Sigma^{2}}\bigg],
\end{equation}
with centroid energy $E_{c}$ given by
\begin{equation}
E_{c} = E\bigg( 1  +\frac{4kT}{m_{\mathrm{e}}c^{2}} - \frac{E_{i}}{m_{\mathrm{e}}c^{2}}\bigg),
\end{equation}
and the standard deviation
\begin{equation}
\Sigma = \Sigma(E_{i}) = E_{i}\bigg[ \frac{2kT}{m_{\mathrm{e}}c^{2}} +
\frac{2}{5}\bigg(\frac{E_{i}}{m_{\mathrm{e}}c^{2}}\bigg) \bigg]^{1/2}.
\end{equation}

This Gaussian approximation was first used in the context of photoionization
models by \cite{ros78PhDT} and \cite{ros78}, who later implemented into the
reflection model {\tt reflion} \citep{ros93}, and its subsequent incarnations
{\tt reflionx} \citep{ros05}, {\tt refbhb} \citep{ros07}, and several other
calculations based on the same code \citep{bal01,bal02,bal04,bal04b,bal05,bal12}.
Likewise, the same approximation was used in calculations using
the {\sc xstar} photoionization code \citep{kal01} to solve the X-ray reflection
problem in hydrostatic atmospheres \citep{nay00,nay01}; and more recently
in the constant density reflection calculations using the \xillver\ code 
\citep{gar10,gar11,gar13a,gar14a,gar16b}.

In all the works mentioned above, the normalization of the probability function
has been chosen such as the integral over the redistribution function gives 
the Klein--Nishina cross section $\sigma_{\mathrm{KN}}(E_{i})$
\begin{equation}
R_\mathrm{G}(E_{i}, E_{f}) = \frac{P(E_{i}, E_{f})\sigma_{\mathrm{KN}}(E_{i})}{\int P(E_{i}, E_{f})dE_{f}},
\end{equation}
which can be explicitly written as
\begin{equation}
\begin{aligned}
\sigma_{\mathrm{KN}}(x) = &
\sigma_{\mathrm{T}}\frac{3}{4}\bigg\{ \frac{1+x}{x^{3}}\bigg[ \frac{2x(1+x)}{1+2x}-\ln(1+2x)
\bigg] \\  
&  +\frac{1}{2x}\ln(1+2x) - \frac{1+3x}{(1+2x)^{2}}\bigg\},
\end{aligned}
\end{equation}
\\
where $x=E/m_{\mathrm{e}}c^{2}$ \citep{kle29}.
As we show next, this normalization needs to be adjusted in order to compare with 
the exact solution.

\subsection{Normalization of the Redistribution Functions}

In order to compare the Gaussian approximation with the exact solution of the 
redistribution function, one needs to make sure that they are both normalized 
in such a way that the integrals $\int R_{\mathrm{G}}(E_{i}, E_{f})dE_{f}$
and $\int R_{\mathrm{E}}(E_{i}, E_{f})dE_{f}$ yield the same total scattering
cross section. At electron temperatures low enough ($T\lessapprox 1\times10^{7}$ K), 
these integrals are equal to Klein--Nishina cross section $\sigma_{\mathrm{KN}}(E_{i})$.

However, at higher temperatures the thermal motion of relativistic electrons in
hot plasma becomes important and needs to be taken into account. The Compton
scattering cross section convolved with a relativistic Maxwellian distribution
of electrons was written by \citet{pou96}:
\vspace{1cm}
\begin{widetext}
\begin{equation}\label{eq:scs}
\begin{aligned}
\sigma_{\mathrm{CS}}(x) = & \frac{3\sigma_{\mathrm{T}}}{16x^{2}\Theta K_{2}(1/\Theta)}
\int^{\infty}_{1} \mathrm{e}^{-\gamma/\Theta}
\bigg\{ \left(x\gamma + \frac{9}{2} + \frac{2\gamma}{x} \right)
\ln\left[\frac{1+2x(\gamma+z)}{1+2x(\gamma-z)}\right]                  
- 2xz + z\left(x-\frac{2}{x}\right)
\ln(1+4x\gamma+4x^{2})                            \\
& + \frac{4x^{2}z(\gamma + x)}{1+4x\gamma+4x^{2}} -
2\int^{x(\gamma+z)}_{x(\gamma-z)}\ln(1+2\xi)\frac{d\xi}{\xi}\bigg\}d{\gamma},
\end{aligned}
\end{equation}
\end{widetext}
where $\Theta=kT/m_{\mathrm{e}}c^{2}$ and $z=\sqrt{\gamma^{2}-1}$.

\begin{figure}[ht]
\centering
\includegraphics[width=\linewidth]{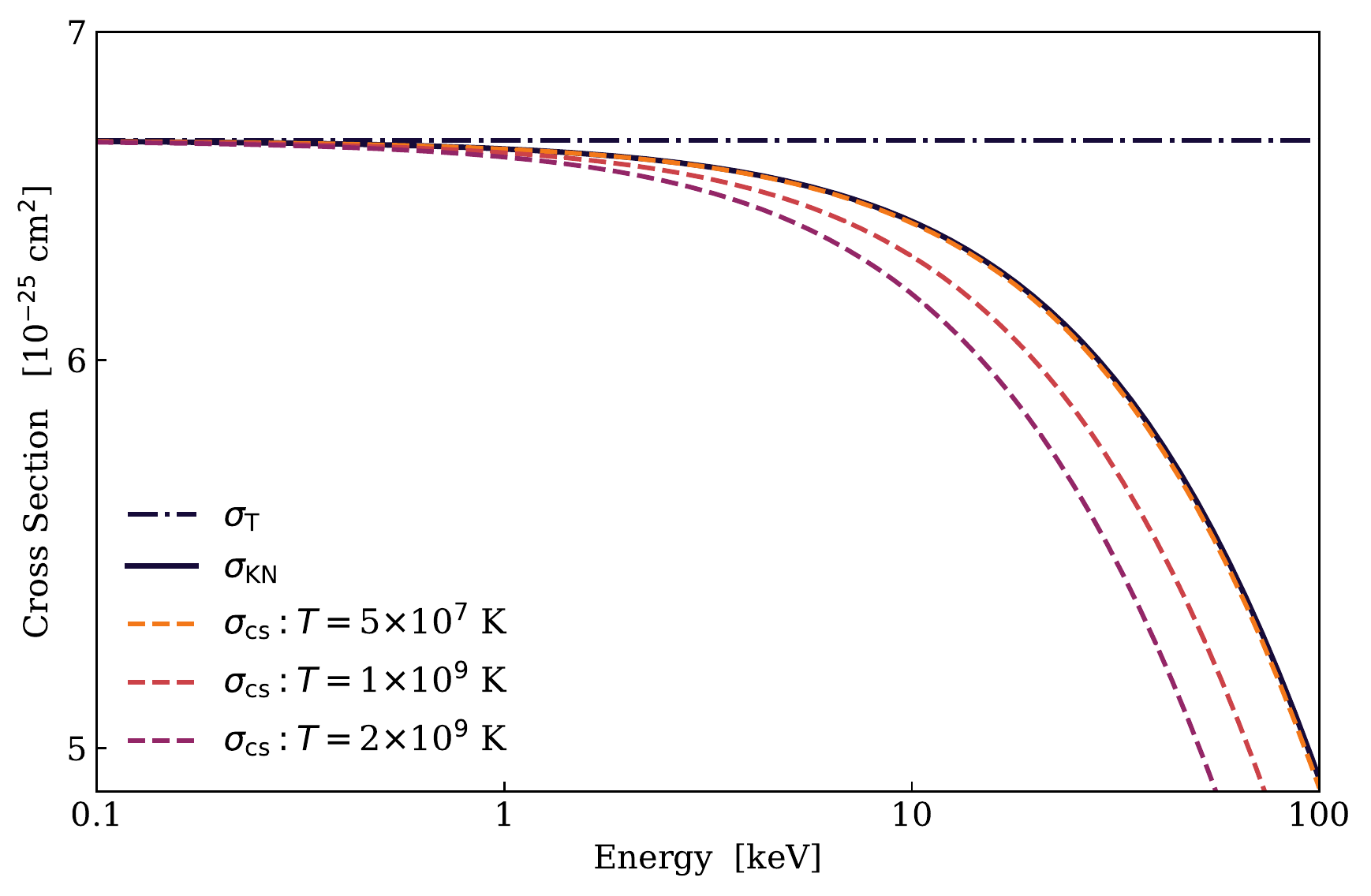}
\caption{
Comparison of the total quantum electrodynamic Klein--Nishina cross section for
Compton scattering of photons off stationary electrons \citep{kle29}, and the
Klein--Nishina cross section convolved with a relativistic Maxwellian
\citep{pou96}. As expected, the largest deviations occur at high electron
temperatures.
}
\label{fig:totalcs}
\end{figure}

We calculated the total cross section in Equation~\ref{eq:scs} using 20-point
Gauss--Legendre quadrature for the internal integral over $\xi$, and 200-point
Gauss--Laguerre quadrature for the integral over $\gamma$.
Figure~\ref{fig:totalcs} shows a comparison of the total cross section averaged
over the relativistic Maxwellian distribution for different electron temperatures,
the classical Thomson, and the Klein--Nishina cross sections.

In order to have a meaningful comparison, we normalize the two redistribution functions
(Exact and Gaussian) to the energy dependent cross section $\sigma_{\mathrm{CS}}$ in the
usual way:
\begin{equation}
R_\mathrm{\{E,G\}}(E_i, E_f) = R_\mathrm{\{E,G\}}(E_i, E_f)
\frac{\sigma_{\mathrm{CS}}(E_f)}{\int R_\mathrm{\{E,G\}}(E_i,E_f)dE_f}.
\end{equation}
We note that the integration over final energies of the exact redistribution
function (from NP93), multiplied by the Thomson scattering cross section, should
itself give the cross section $\sigma_{\mathrm{CS}}(E)$. But, accounting for
the finiteness of the grids used later for radiative transfer calculation, and
thus, the errors in numerical integration over angles and $\gamma$, the
normalization is still required.

%......................................................................................
%......................................................................................

\begin{figure*}[ht!]
\centering
\includegraphics[width=\linewidth]{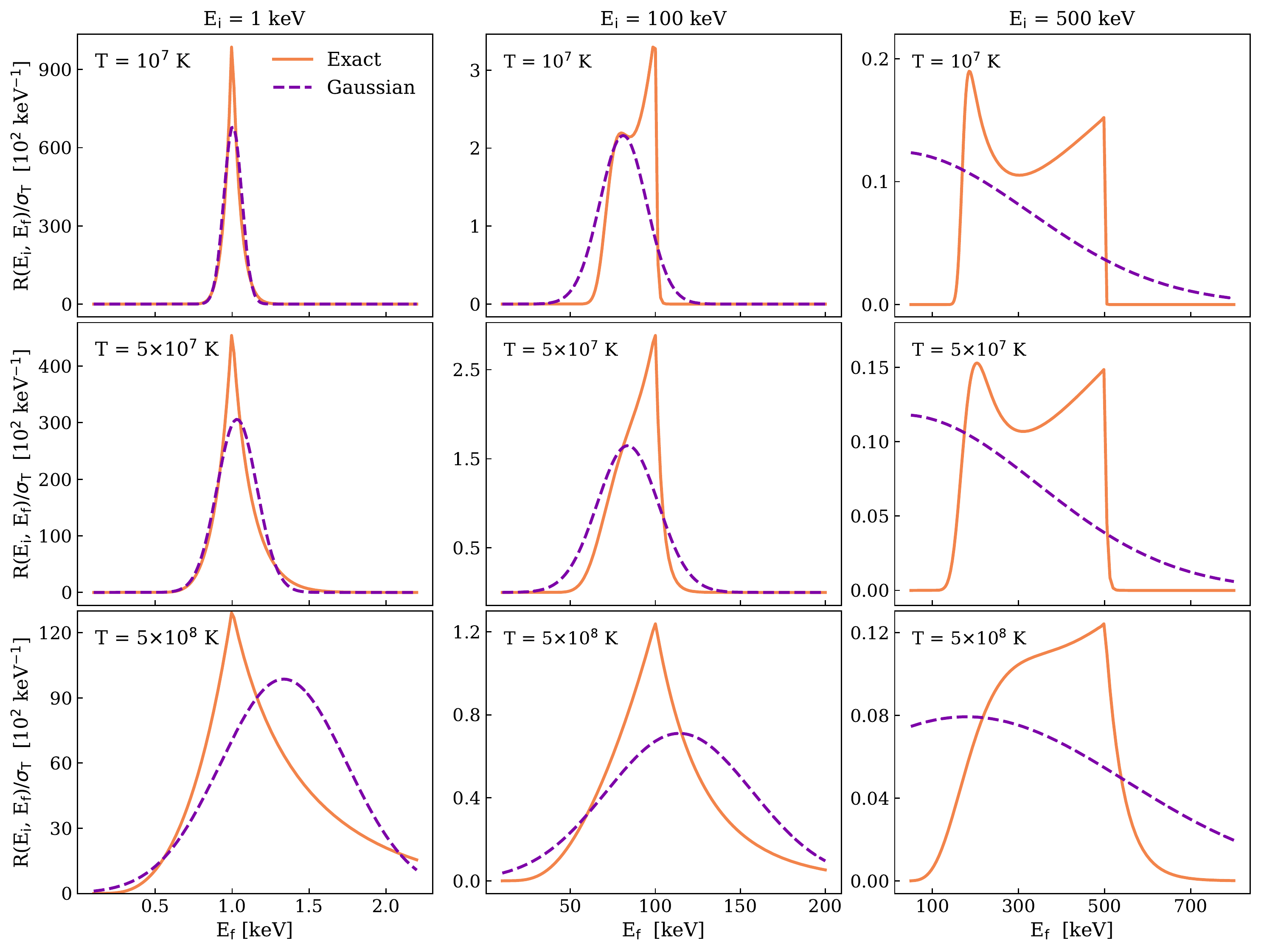}
\caption{
Comparison of the Gaussian approximation for the redistribution of photons after
electron scattering events (adopted in several reflection codes such as \reflionx\
and \xillver), and the fully relativistic solution by \cite{nag93}.
The two solutions agree well at low energies, but at high energies the
discrepancy is dramatic. A similar trend is also observed with respect to the
electron temperature.
}
\label{fig:rf_all}
\end{figure*}

Figure~\ref{fig:rf_all} shows a comparison of the redistribution functions
computed with the Gaussian approximation and the exact solution using the above
normalization. We used up to 3000-point Gauss-Legendre quadrature to integrate
the redistribution function Eq.~\ref{redex} over the scattering angle. The
integral over $\gamma$ is computed using 32-point Gauss-Laguerre quadrature
after setting the new variable $u=(\gamma-\gamma_{*})/\Theta$, as discussed in
\citet{mad17}. The modified Bessel function has to be treated with care, thus
at the limiting cases we used asymptotic expansions, given by
\citet{abram1988}.  We show comparisons for different initial photon energies
and electron temperatures. From this comparison is evident that the Gaussian
approximation works much better at low photon energies and electron
temperatures.  

For an initial photon of 1\,keV, the exact solution is symmetric and thus the
Gaussian approximation agrees fairly well, except at very high temperatures
($T\gtrsim 10^8$\,K) where the peaks of the distributions are shifted.  At
higher energies, as the photon approaches $m_e c^2$, the exact
solution becomes double-peaked and more asymmetric, making the differences with
the Gaussian approximation more noticeable.  When the energy of the photons
is much larger than the kinetic energy of the electrons, the solution
approaches the case of a single scattering out of electrons at rest. In this
limit, the shape of the redistribution function is symmetric and double peaked, as shown
by \cite{lig81}. The solution peaks at the initial photon energy, and at the
energy of maximum shift, given by twice the Compton wavelength $\lambda_c$.  In
this regime, the Gaussian solution becomes extremely broad, failing to
accurately represent the correct redistribution function due to Compton
scattering. Nevertheless, up to $\sim 100$\,keV, the agreement between the two
solutions is acceptable.

%......................................................................................
%......................................................................................

%----------------------------------------------------------------------------------
\section{Results}\label{sec:results}

\subsection{Convolution of an Input Spectrum with the Redistribution Functions}

We have conducted calculations to estimate the discrepancies between the Gaussian
approximation and the exact solution of the redistribution function discussed in 
the previous Section might affect the resulting spectrum after several scattering events.
This is done by convolving an input spectrum
with each one of two the redistribution functions mentioned above in the following way:
\begin{equation}\label{eq:jprim}
J^{\rm n}_{\rm c}(E_{i}) =
\frac{1}{\sigma_{\mathrm{CS}}(E_{i})}\int^{E_{\mathrm{max}}}_{E_{\mathrm{min}}}dE_{f}J^{\mathrm{n}-1}_{\rm c}(E_{f})R(E_{f},
E_{i}),
\end{equation}
where $n$ is the number of scattering events. For $n=1$, $J^{\mathrm{n}-1}_{\rm c}(E)$ 
represents the input spectral distribution of photons propagating through scattering 
medium.
Notice that in the scattering integral of Equation~\ref{eq:jprim} the inverse
redistribution function $R(E_f, E_i)$ needs to be used
(rather than $R(E_i, E_f)$). This is because the integrand
requires the probability that a photon from any other energy $E_f$ is
scattered into the current energy of interest $E_i$.  Importantly, this similar 
convolution procedure required during the solution of the transfer
equation in reflection codes such as \xillver. We shall come back to the
discussion of these calculations in the following Sections.

\begin{figure*}[ht]
\centering
\includegraphics[width=0.95\linewidth]{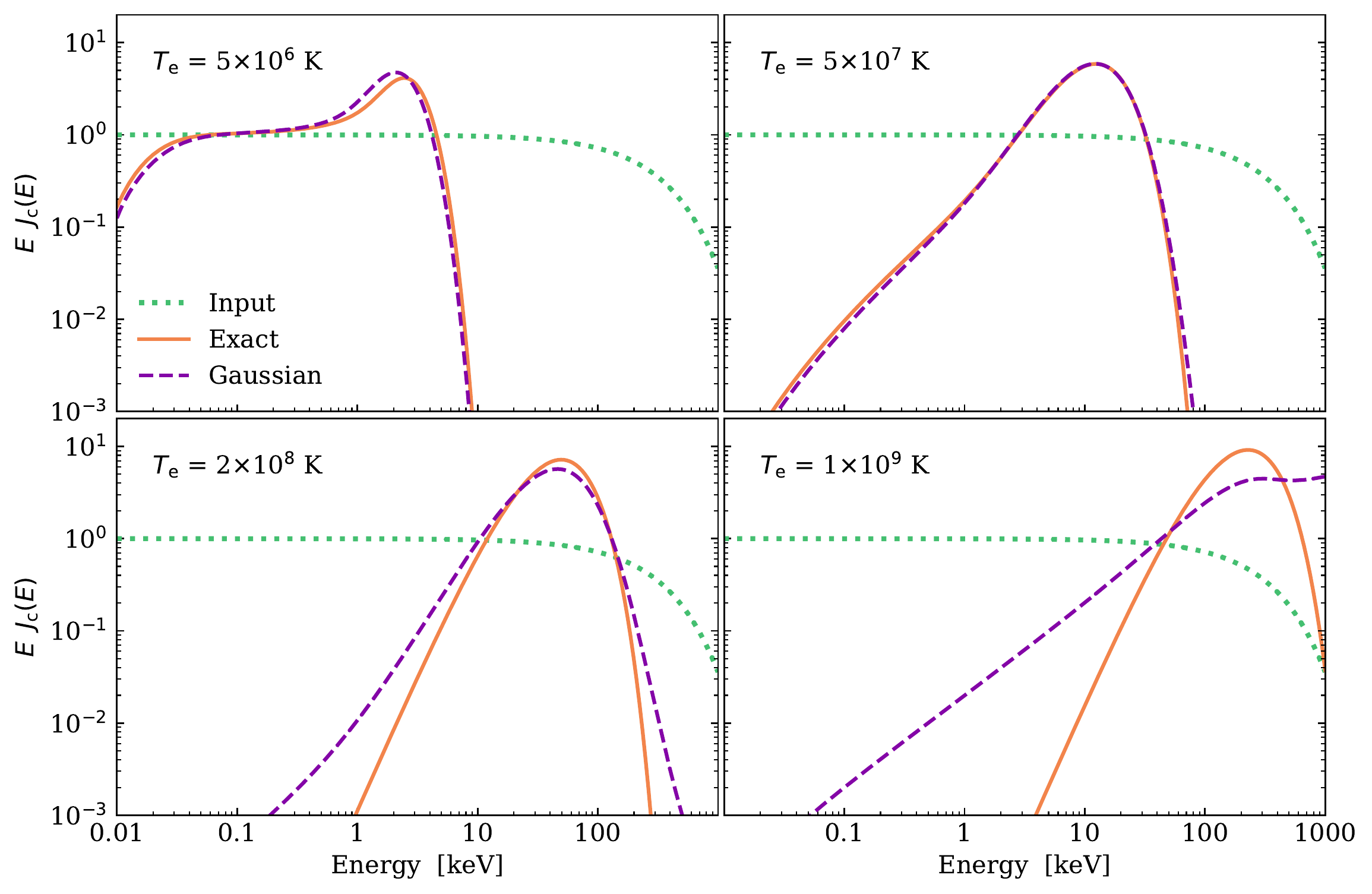}
\caption{
The result of 300 convolutions (Eq.~\ref{eq:jprim}) for the input power law continuum with 
high-energy exponential cutoff for the Gaussian approximation (dashed purple line) and 
the exact solution given by NP93 (solid yellow line). The input spectrum is shown by 
dotted green line. 
}
\label{fig:conv_tdiff}
\end{figure*}

Figure~\ref{fig:conv_tdiff} shows the results for an input power-law with a
high-energy cutoff in the form $F(E) \propto E^{-(\Gamma-1)}\exp{(-E/E_\mathrm{fold})}$,
with $E_\mathrm{fold}=300$\,keV, $\Gamma=2$ (similar to the canonical input spectra used
in the reflection calculations). The convolution is done 300 times to account
for multiple scatterings. At a relatively low temperature ($T=5\times10^{6-7}$\,K),
the result calculated using the Gaussian approximation is very close to the one 
obtained with fully relativistic redistribution function. However, at higher 
temperatures ($T=10^9$\,K), the discrepancies are much more dramatic, with strong 
deviations from the exact solution in the entire energy range.

\begin{figure*}[ht]
\centering
\includegraphics[width=0.9\linewidth]{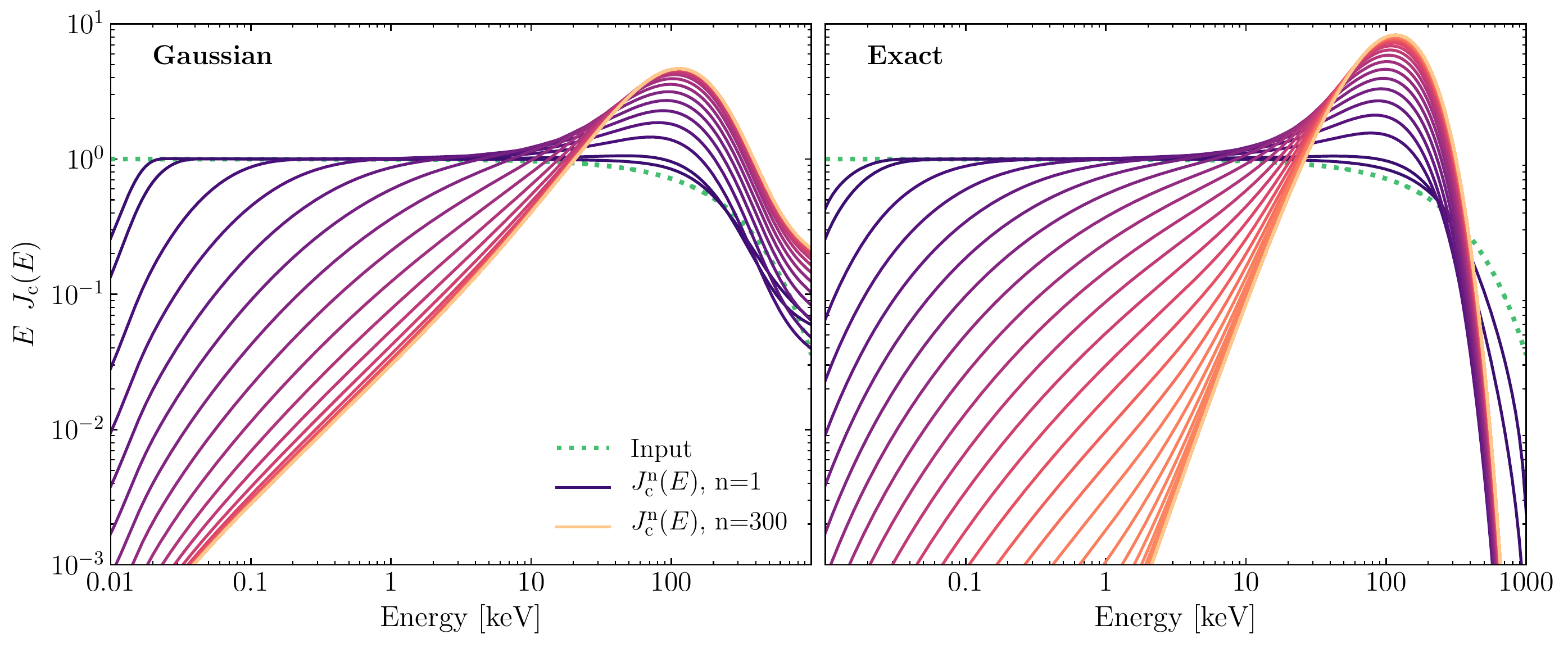}
\caption{
The gradual changes in resulted spectrum during the performance of the multiple convolutions.
$J^{\mathrm{n}}_{\rm c}(E)$ is shown for every 4th iteration for clarity. The left-hand panel
shows the results from the Gaussian approximation, while the right-hand panel shows those from
the NP93 exact solution. The same input spectrum as for Figure~\ref{fig:conv_tdiff} was used. 
}
\label{fig:conv_iter}
\end{figure*}

After many iterations the synthetic spectrum reaches some saturation, approaching a
Wien distribution peaking at $E\approx 2.8 kT_{\rm e}$, which is expected for the 
case of the saturated Comptonization \citep{sutit1980,hua95}. Figure~\ref{fig:conv_iter} 
shows the gradual changes in the spectra while approaching some equilibrium condition 
for both redistribution functions discussed above.

%......................................................................................
%......................................................................................

%......................................................................................
%......................................................................................

\subsection{Radiative Transfer Calculations}

We now test the effects of the different solutions for the redistribution of
photons due to Compton scattering in the final solution of the X-ray spectrum
reflected from the surface of an accretion disk. To this end we make use of 
the routines in our reflection code \xillver. This model assumes a plane-parallel
geometry for a constant density slab of a few Thomson depths (typically 
$\tau_\mathrm{T}\sim 10$). The radiative transfer is solved using the Feautrier
method with two boundary conditions for the incident field at the top and bottom
of the atmosphere. A detailed description of the numerical methods employed in
the \xillver\ code can be found in our  previous publications \citep{gar10,gar13a,gar14a}.

Until now, \xillver\ has made use of the Gaussian approximation to account for
the redistribution of photons due to Compton scattering. We have now modified these
routines to implement the exact solution for the redistribution function by
NP93, including relativistic corrections to the total Compton cross section
\citep{pou96}, as described in Section~\ref{sec:numeric}.

\subsubsection{Pure Scattering Case}

In order to test the effects of the new redistribution function, 
we start with the simple case in which electron scattering is the only source of 
opacity, and no thermal emission coefficient, i.e., $\sigma_a=0$. Furthermore, 
thermal equilibrium is not imposed, such that the gas temperature is kept fixed
at a given value. In this configuration, the
source function for pure scattering in an isothermal atmosphere is represented by
the double integral:
\begin{equation}
S(E) = J_c(E) = \frac{1}{2\sigma_\mathrm{CS}} \int{dE' R(E',E)\int_{-1}^{1}
d\mu I(E',\mu)}.
\end{equation}
Based on the comparisons discussed above, we expect that the largest differences
appear at high temperatures and photon energies. Thus, calculations were carried out
assuming an isothermal atmosphere with constant number density of $10^{15}$\,cm$^{-3}$.
The illumination from above was assumed to be a
relatively weak power-law with $\Gamma=2$ at 45\deg\ incidence, and a much 
stronger isotropic black body radiation field with $kT=0.35$\,keV entering from below at 
$\tau_\mathrm{T}=10$. This particular choice of illumination is somewhat arbitrary,
but it serves to test the effects of the Comptonization in the case of a radiation
field of a blackbody type. Such a setup can resemble the case of a bright black hole
binary system in the soft state, during which the accretion disk becomes luminous
and dominates the emission of the X-ray spectrum \citep[e.g.][]{mcc06}.

\begin{figure}[ht]
\centering
\includegraphics[width=\linewidth]{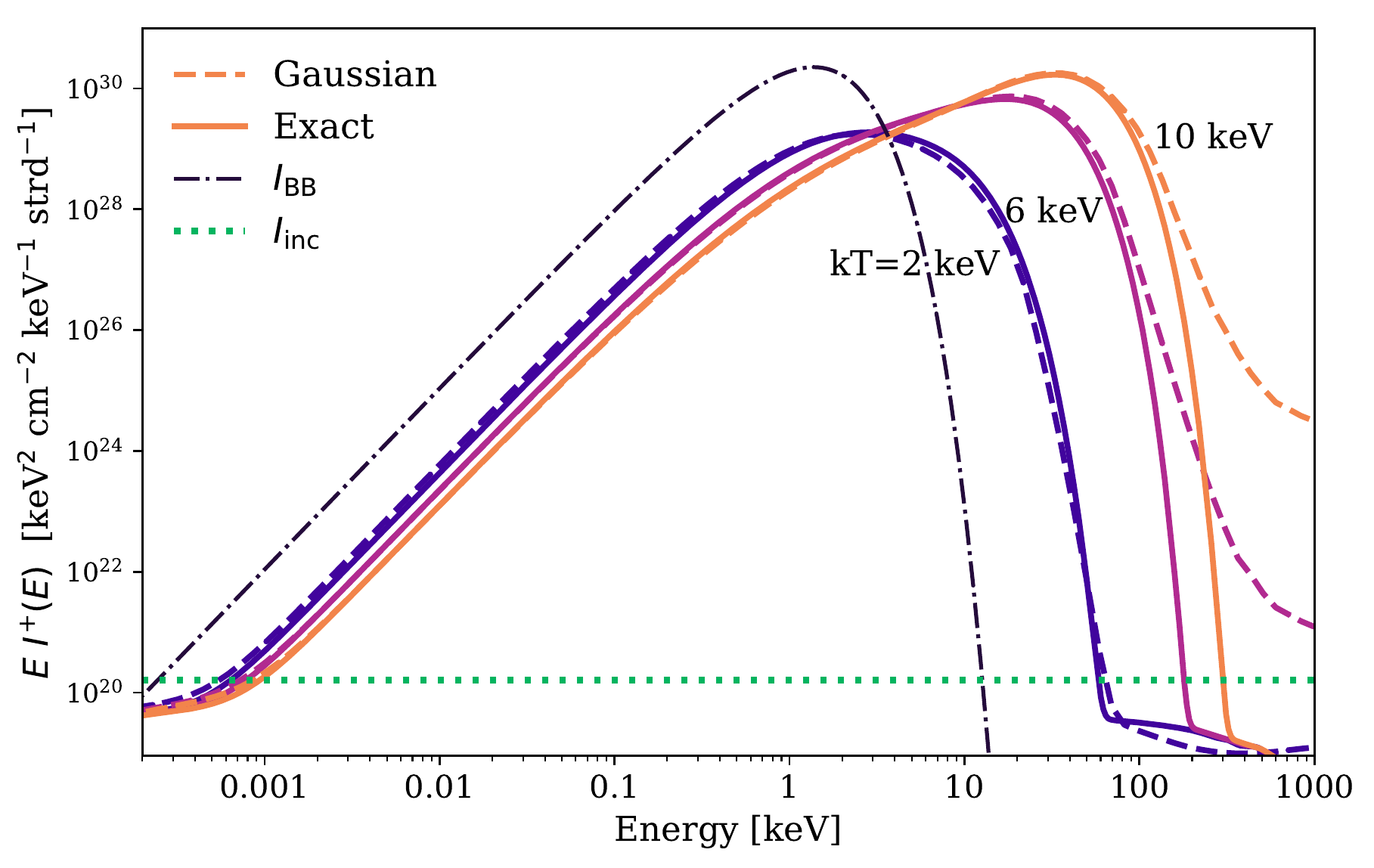}
\caption{
Radiative transfer calculations of X-ray reflection spectrum from a slab
including only Compton scattering (i.e., no atomic contributions, no pair
production). The slab is illuminated by a power-law at the top, and by strong
thermal emission at the bottom. Shown are three cases for the disk temperature,
comparing the exact solution with the previous Gaussian approximation.
}
\label{fig:purescat}
\end{figure}

Comparisons of the calculations done with the usual Gaussian approximation and with
exact solution with the \xillver\ routines are shown in Figure~\ref{fig:purescat},
for three different electron temperatures, $kT_e=2, 6$, and $10$\,keV. 
Comptonization in this hot medium produces a significant modification of the original
black body field, which becomes more severe at higher electron temperatures. For
the coldest case, the Gaussian solution approximates well to the exact. However,
for the other two cases, large discrepancies are obvious for photon energies above
$\sim 100$\,keV, where the Gaussian approximation underestimates the amount of photons
that get scattered to lower energies. 

%......................................................................................
%......................................................................................

\subsubsection{Scattering and True Absorption at Fixed Temperature}

Next, we have carried out calculations of reflected spectra now including all astrophysically
relevant atomic transitions (assuming Solar abundances), for a slab at constant density
($n_e=10^{15}$\,cm$^{-3}$) and constant 
temperature ($T\sim 1$\,keV), for 3
different ionization parameter $\xi = 10, 10^2, 10^3$\,erg\,cm\,s$^{-1}$. In this case,
we follow a more standard setup for the reflection calculation: the slab is illuminated
on the top by a power-law spectrum with $\Gamma=2$ and a high-energy cutoff at 300\,keV.
Contrary to the pure scattering models presented in the previous Section, in this case
no illumination from below is considered. This particular setup resembles the case of a
black hole binary in the hard state, during which the disk emission is faint, while the 
X-ray spectrum is dominated by a non-thermal power-law like emission \citep[e.g.][]{rem06,mcc13}; or
the case of reflection in accretion disks around supermassive black holes the AGN of
Seyfert galaxies \citep[e.g.][]{rey13}. This is a common and canonical setup for reflection
model calculations \citep[e.g.][]{ros05,gar10,gar13a} As before, Compton scattering inside
the slab was calculated with the two redistribution functions
(Gaussian and exact), keeping the rest of the model unchanged. 

\begin{figure}[ht]
\centering
\includegraphics[width=\linewidth]{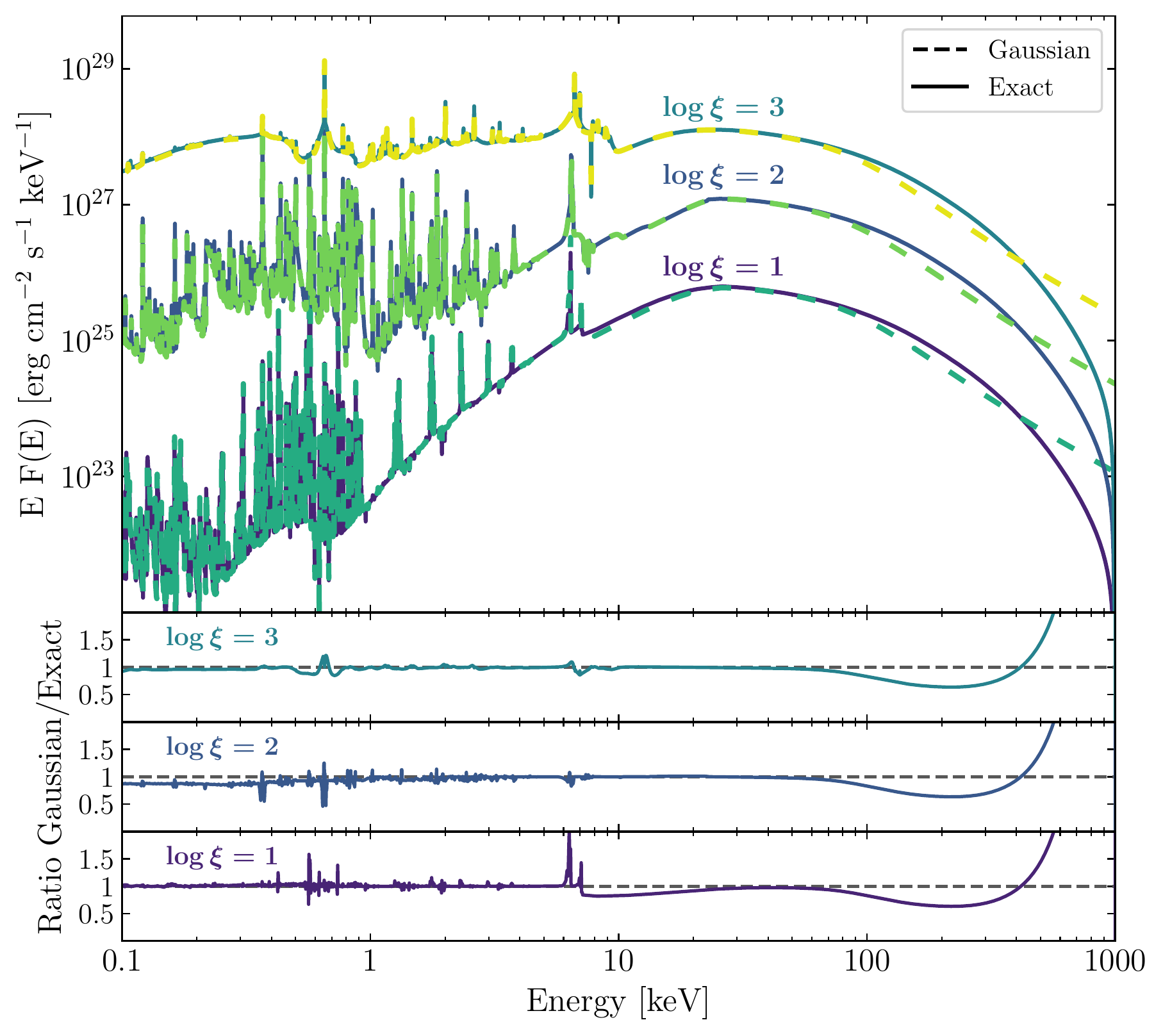}
\caption{
({\it Top}) Angle-averaged reflected spectra calculated with the \xillver\ code for a 
constant density ($n_e=10^{15}$\,cm$^{-3}$) and isothermal ($T\sim1$\,keV) atmosphere,
including both scattering and atomic opacities. The slab is illuminated at the top 
by a power-law spectrum with $\Gamma=2$ and a high-energy cutoff at $300$\,keV.
Results for 3 different ionization parameters ($\xi = 4\pi F_x/n_e = 10, 10^2, 
10^3$\,erg\,cm\,s$^{-1}$) are presented, using both the Gaussian approximation (dashed curves)
and th exact solution (solid curves) by NP93. ({\it Bottom}) Ratio of the Gaussian to the Exact
spectra for each value of the ionization parameter, as indicated.
}
\label{fig:xillver_comp}
\end{figure}

Figure~\ref{fig:xillver_comp} shows the comparison of the reflected spectra after
200 $\Lambda$-iterations for the radiative transfer solution. As expected, we find
that the largest differences are at high energies (above $\sim100$\,keV), with the
exact solution producing a more curved reflection spectrum as compared to the 
Gaussian approximation. The rest of the reflection spectrum appears mostly unaffected
for this particular configuration of parameters. However, a more detailed inspection 
of the ratio of the Gaussian to the Exact solution spectra (Fig.~\ref{fig:xillver_comp},
bottom panels), shows differences in some of the lines profiles. Specifically, the
iron K-shell emission lines appear stronger in the Gaussian solution, indicating that
the amount of Comptonization in the lines is underestimated. This difference is stronger
at low ionization, when the Fe K emission is dominated by the narrower complex of lines
at $\sim6.4$\,keV (K$\alpha$) and $\sim7.1$\,keV (K$\beta$). Some smaller differences 
are also present at lower energies, in particular for the oxygen Ly-$\alpha$ emission 
at $\sim0.8$\,keV.

\subsubsection{Complete Reflection Calculations}\label{sec:complete}

The ultimate goal is to implement the new and exact solution for the
Comptonization into a full reflection calculation. For this, we produced a set
of \xillver\ models in which we now allow the code to solve the energy and
ionization balance equations at each point in the atmosphere. With these models
we aim to test the possible secondary effects in the solution due to the change
in the energy budget of the radiation field as a consequence of the improved
redistribution function. Based on all the previous tests, we expect the largest
differences to appear when the irradiated atmosphere reaches the highest
temperatures. Thus, we have run models with a larger ionization parameter of
$\xi=3\times10^3$\,erg\,cm\,s$^{-1}$, while keeping the same gas density of
$n_e=10^{15}$\,cm$^{-3}$ (which implies an increase of the net flux incident at
the top); and with a harder slope for the illuminating spectrum
$\Gamma=1.6$ (which increases the number of photons at high energies).  To enhance
the strength of the Fe K emission profile, we set the iron abundance to five times its
Solar value ($A_\mathrm{Fe}$), which is also in line with the values required
to fit reflection in many accreting black holes \citep{gar18a}.

\begin{figure*}[ht]
\centering
\includegraphics[width=\linewidth]{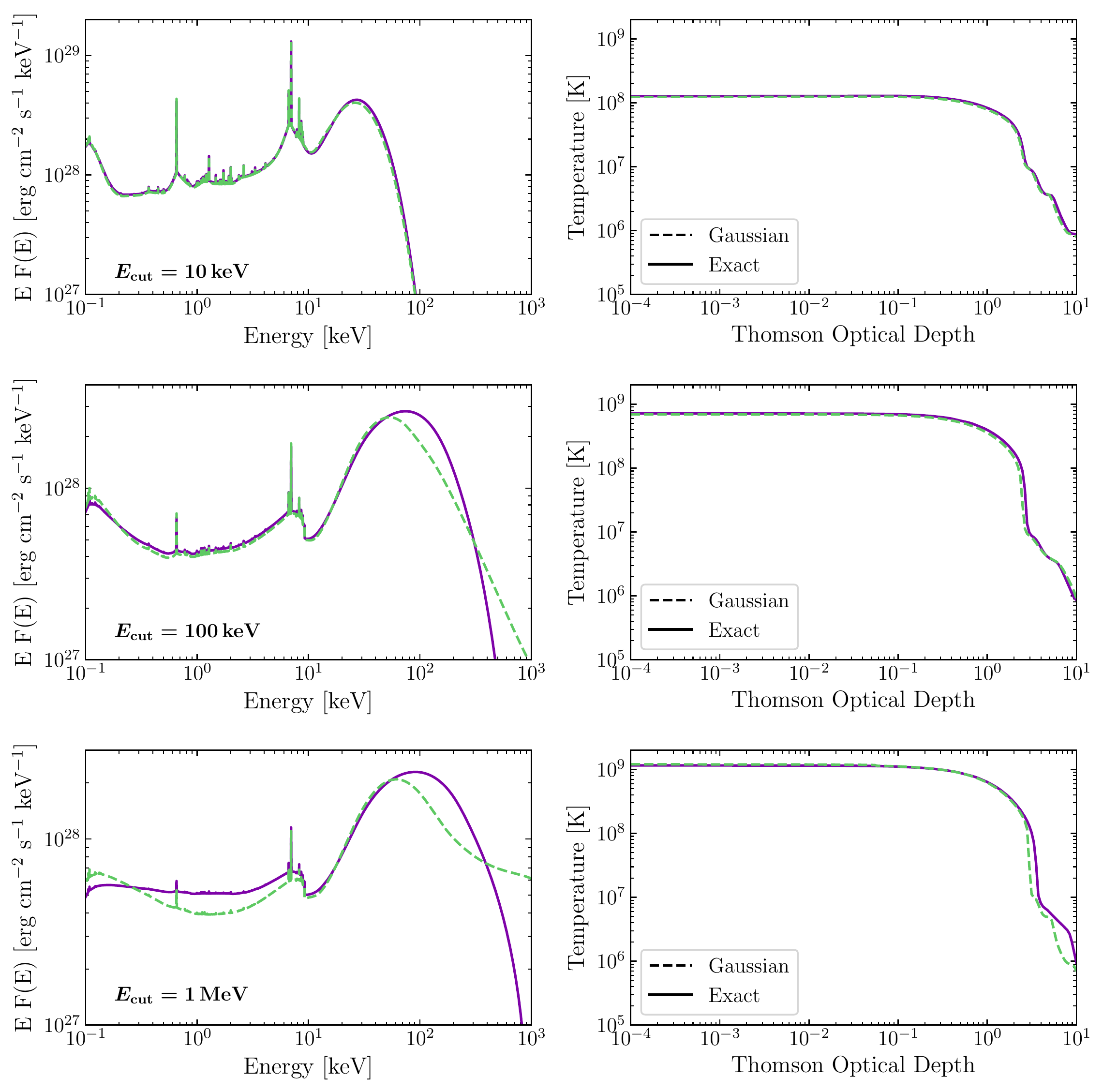}
\caption{
Full \xillver\ reflection calculations for different illuminating spectra, including
the solution of the ionization balance and energy equations. The left panels show the 
angle-averaged reflected spectrum, while the right panels show the corresponding temperature
profiles in the atmosphere. The irradiation is assumed to a power law with $\Gamma=1.4$,
and 3 different high-energy cutoffs ($10, 10^2, 10^3$\,keV), as indicated. The ionization
parameter is fixed at $\xi=3\times 10^3$\,erg\,cm$^2$\,s$^{-1}$, and the abundance of
iron at 5 times the solar value. Solutions using the Gaussian (dashed) and Exact (solid)
redistribution functions are compared.
}
\label{fig:xillver}
\end{figure*}

Three main calculations were done only varying the high-energy cutoff of the illuminating
power-law, which was set at $E_\mathrm{cut}=10, 10^2$, and $10^3$\,keV. All other parameters
were kept at the values quoted above.
The main results of the full \xillver\ calculations are presented in Figure~\ref{fig:xillver},
which includes both the reflected spectra at the surface, and the full temperature 
profile in the vertical direction of the atmosphere.
As before, we show the comparisons
of the same calculation carried out with the Gaussian and the exact redistribution functions
for Compton scattering.  

The largest and most obvious differences in the reflected spectrum are seen at
high energies (above $\sim 20$\,keV), where the Compton kernel acts most
strongly. However, for the lowest value of the high-energy cutoff
($E_\mathrm{cut}=10$\,keV) the results look almost identical, with very minor
difference in the temperature profile at the largest optical depth. This is
likely due to the lack of high-energy photons in the illumination, and the fact
that the temperature never reaches extreme values. The strongest atomic
features in these models are due to O ($\sim 0.8$\,keV) and Fe ($\sim
6.7$\,keV) K-shell emission. 

We notice that the peak of the emission line does
not change among calculations with the different redistribution functions. This
is because those line-core photons are mostly emitted close to the surface, and
thus they suffer little to none scatterings on their way to the observer. The
temperature profiles show relatively small variations between the two
approximations. Only in the hottest case ($E_\mathrm{cut}=1$\,MeV) we see a
noticeable difference in the deepest regions of the slab, likely because the
radiation field has been modified sufficiently such that the temperature solution
is affected. 

Evidently, the largest differences in the spectra are observed in
the region of the Compton hump, and they are most marked for models with
highest cutoff energy. This is expected, as these are the models with enough
photons in the energy range where Comptonization produces the strongest
redistribution, but also because the higher $E_\mathrm{cut}$ the higher
the overall temperature of the atmosphere.

In fact, the model with $E_\mathrm{cut}=1$\,MeV shows strong departures from
the Gaussian approximation at all energies, even below 1\,keV. One reason for
the changes at soft energies is the different solution of the ionization
balance, as clearly evidenced by the discrepant temperature profiles at large
depths. However, a more important effect is introduced by the Comptonization
solution, given that the gas temperature is close to $10^9$\,K in a large
portion of the slab at the upper layers ($\tau_\mathrm{T}\sim10^{-4}-1$). As
discussed in previous sections, in this high temperature regime the
redistribution function based on the Gaussian approximation fails most
dramatically when compared to the exact solution (e.g.,
Figure~\ref{fig:rf_all}). When the temperature of the atmosphere reaches such
high values, the Gaussian solution of the Comptonization produces inaccurate
results at all photon energies.

%......................................................................................
%......................................................................................

%......................................................................................
%......................................................................................

%----------------------------------------------------------------------------------

\section{Discussion and Conclusions}\label{sec:disc}

In this paper we have presented a detailed discussion of the thermal Comptonization
process in optically thick media. We have concentrated our results to the context
of the reprocessed high-energy radiation in accretion disks around compact objects. 
Still, the approximations presented here are useful to any other problems in which
accurate calculations of the Comptonized spectrum are required.

Given the intrinsic complexity of the X-ray reflection calculations,
traditional models have made use of a simplified Gaussian approximation to
describe the redistribution of photons in the X-ray band after suffering many
scatterings in an optically-thick slab.  We have shown new calculations that
allow to adopt a much more accurate solution for the Comptonization,
taking into account the most relevant physics (i.e., quantum electrodynamical
and relativistic corrections to the classical Thomson cross section).
Therefore, the NP93 solution, which is the solution implemented in the present 
work, can be considered as {\it exact} within the precision of the numerical
integration of the Maxwellian distribution. 

Comparisons between the previously used Gaussian approximation and the exact
NP93 solution reveal that the major discrepancies occur at either very high
energies (typically above $\sim 100$\,keV), where electron scattering becomes
most important; and/or when the gas temperature approaches $\sim 10^9$\,K
(regardless of the photon energy), a regime in which the thermal motions of the
electrons become relativistic causing a decrease of the Klein--Nishina cross
section at energies above $\sim 1$\,keV (Figure~\ref{fig:totalcs}).  We emphasize
that such high temperatures do not represent the most common conditions of
accreting sources, in which coronal temperatures of hundreds of keV are typically
observed \citep[e.g.][]{fab15,fab17}.

Meanwhile, the changes of the spectral shape in the
Compton hump band ($\sim 20-40$\,keV) observed in the reflection spectrum can potentially affect the coronal
temperature derived from spectral fits. Coronal temperatures are typically estimated by measuring the
cutoff at high energy of the continuum and reflection spectra \citep[e.g.][]{gar15a,kar17,bui19}. The reflection
spectra using the new Comptonization solution shows a sharper cutoff than the
one produced with the Gaussian solution, but depending on the parameter it
might also appear at higher energies. A proper assessment of the differences in the
recovered coronal temperatures from the application of these models to observational
data requires the calculation of a complete grid of models covering a wide range
of parameters. Such a effort is outside the scope of the present paper. However, based
on the results presented in Section~\ref{sec:complete},  we expect these
differences to be relatively small (of the order of tens of keV or less). This is
because the largest differences seem to appear when the cutoff energy (or coronal
temperature) is relatively high (close to 1\,MeV), while astrophysical black holes
are expected (and typically observed) to have coronal temperatures of hundreds of keV
or less. 

Previous works on spectral models that have made use of the
reflection tables produced with our \xillver\ code have pointed out
discrepancies found in the Compton hump at high energies, when comparing the
reflected spectrum with that produced with a more accurate Monte Carlo
calculation. This has motivated the use of numerical artifacts in these models,
in order to try to correct for the discrepancy.  Such is the case of the {\tt
xilconv} model, a modification of the {\tt rfxconv} model \citep{kol11}, and
first described by \cite{don06}. This model uses the \xillver\ spectra below
14\,keV and the Compton reflection code {\tt pexrav} by \cite{mag95} for higher energies. The
merging of these two products requires a somewhat convoluted procedure, which
is likely to reduce the self-consistency of the model.  A similar approach is
followed in the {\tt reflkerr} model \citep{nie19}, a more recent relativistic
reflection code which also uses the \xillver\ spectra for energies below
$\sim20$\,keV, and the {\tt ireflect} model \citep[which is a generalization of the
{\tt pexrav} model][]{mag95} at high energies.
The new calculations presented here include the correct photon
redistribution due to thermal Comptonization in the reflection calculations
will remove the necessity to modified the \xillver\ spectra. 

%......................................................................................
\begin{figure}[ht!]
\centering
\includegraphics[width=\linewidth]{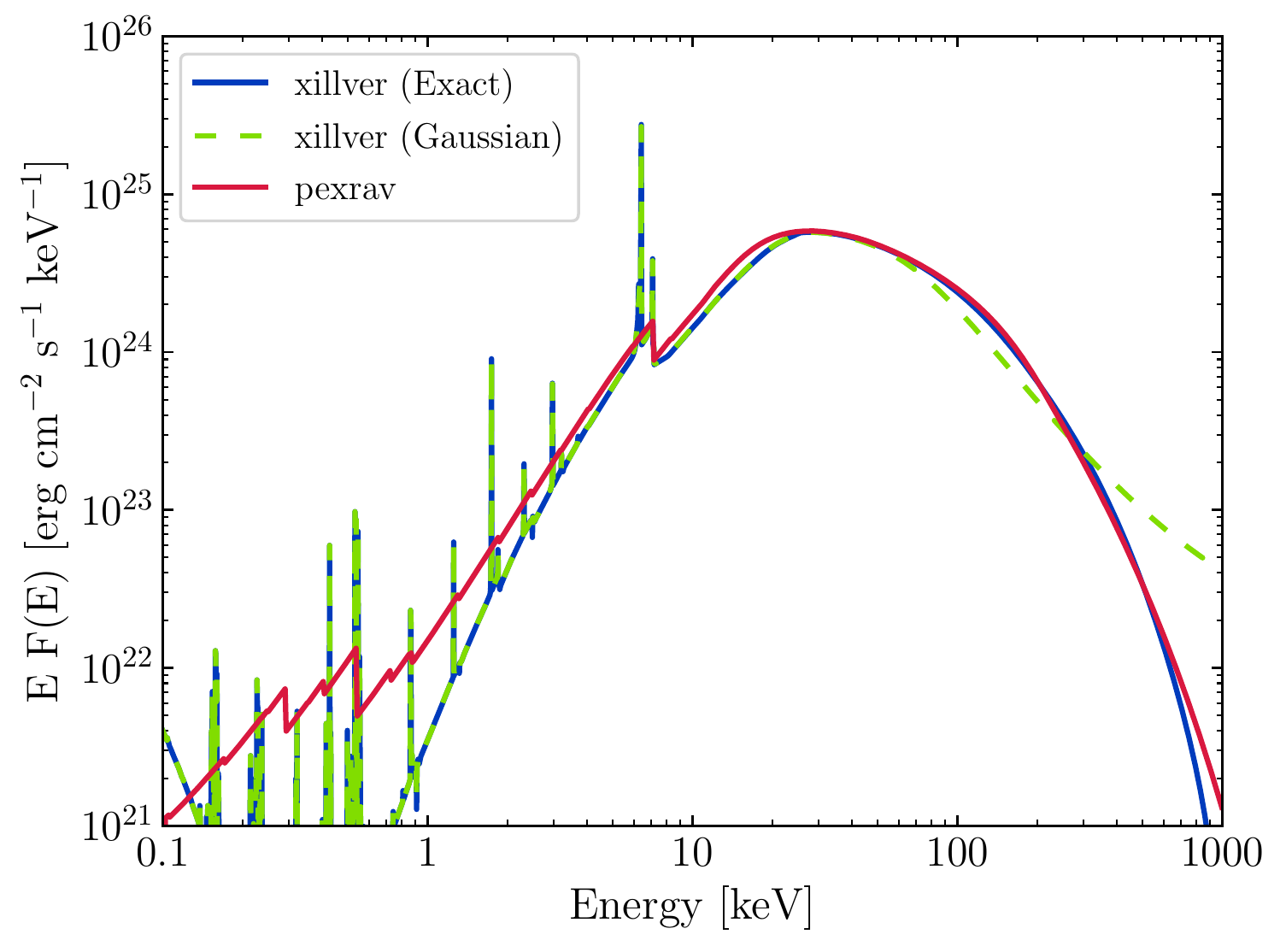}
\caption{
Comparison of the reflected angle-averaged spectra produced with \xillver\
using the Gaussian approximation, the exact solution by NP93, and the
calculation using the {\tt pexrav} code, based on Monte Carlo calculations
\citep{mag95}. The parameters used are $\Gamma=2$, $E_\mathrm{cut}=300$\,keV,
and solar abundances for iron. The \xillver\ spectra were produced for
$\log(\xi/$erg\,cm$^2$\,s$^{-1})=0$.
}
\label{fig:pexrav}
\end{figure}
%......................................................................................

The dramatic improvement brought by the new Comptonization solution to the
\xillver\ calculations is shown in Figure~\ref{fig:pexrav}, where we compare
the angle-average reflected spectrum generated with \xillver\ using both the
Gaussian and the NP93 exact solutions for Comptonization, together with the
calculations using the {\tt pexrav} model. Notice that the latter is a
calculation for a neutral gas (thus neglecting ionization balance), assuming
electrons at rest (thus no temperature dependence in the Comptonization).
Furthermore, the {\tt pexrav} model does not include line emission, but this is
unimportant for the present comparison. Nevertheless, the {\tt pexrav} model is
considered the gold-standard for Compton down scattering.  It is obvious that
the \xillver\ spectrum produced with the NP93 solution for Comptonization
agrees very closely with that from {\tt pexrav} at energies above
$\sim30$\,keV, where Compton scattering dominates the gas opacity.  The small
differences below $\sim 30$\,keV are due the different iron photoelectric
opacities implemented in the two codes. At higher energies both calculations
agree very closely, except near 1\,MeV, where a small divergence is seen. This
is likely due to the fact that this is the upper limit of our calculations,
which could affect our convolution. Moreover, the {\tt pexrav} model is constructed
using fits to detailed Monte Carlo calculations, and thus it could be prone to small
numerical errors.

Importantly, the inclusion of an accurate description of the Comptonization in the
reflection calculations has allowed us to verify and validate the results from
spectral fits that made use of our reflection models. Based on the results
presented here, we are now confident in that the limitations of the Gaussian
approximation used in standard modeling only manifests in extreme cases of
very high temperatures, which will likely have a minor impact in the overall
bread of spectral fits of accreting compact objects published to date.
Nevertheless, future releases of our \xillver\ tables of reflection spectra
will be computed implementing the {\it exact} NP93 solution described here, in
an attempt to provide a yet more accurate prediction of the reprocessed X-ray
spectrum in accretion disks. The implementation of these new models will be
identical to the current version, with no additional parameters.
The present version of the codes used in this paper
for the calculation of the redistribution Compton kernel are made publicly available
on GitHub\footnote{\texttt{driveSRF} codebase:\url{https://github.com/jajgarcia/exact_Compton}}.

%----------------------------------------------------------------------------------
%\subsection{Simulations with HEX-P}

%==================================================================================
%
%
\acknowledgments 
This work was been partially supported under NASA No. NNG08FD60C.  J.A.G.
acknowledges support from NASA ATP grant No. 80NSSC20K0540 and from the Alexander
von Humboldt Foundation. E.K.S and J.W. have been supported by DFG grant WI 
1860/11-1.  A.R., J.M., and A.M. were supported by grants No.
2015/17/B/ST9/03422 and 2015/18/M/ST9/00541 from the Polish National Science
Center.

\vspace{5mm}
\software{
{\sc xillver} \citep{gar10,gar13a}, {\sc Matplotlib} \citep[version 3.1.3,][]{hun07},
{\sc NumPy} \citep[version 1.18.1,][]{oli06}.
}
%
%
%==============================================================================
%
%
\bibliographystyle{aasjournal}
\bibliography{my-references}

\begin{thebibliography}{}
\expandafter\ifx\csname natexlab\endcsname\relax\def\natexlab#1{#1}\fi
\providecommand{\url}[1]{\href{#1}{#1}}
\providecommand{\dodoi}[1]{doi:~\href{http://doi.org/#1}{\nolinkurl{#1}}}
\providecommand{\doeprint}[1]{\href{http://ascl.net/#1}{\nolinkurl{http://ascl.net/#1}}}
\providecommand{\doarXiv}[1]{\href{https://arxiv.org/abs/#1}{\nolinkurl{https://arxiv.org/abs/#1}}}

\bibitem[{{Abramowitz} {et~al.}(1988){Abramowitz}, {Stegun}, \&
  {Romer}}]{abram1988}
{Abramowitz}, M., {Stegun}, I.~A., \& {Romer}, R.~H. 1988, American Journal of
  Physics, 56, 958, \dodoi{10.1119/1.15378}

\bibitem[{{Aharonian} \& {Atoyan}(1981)}]{ahar1981}
{Aharonian}, F.~A., \& {Atoyan}, A.~M. 1981, \apss, 79, 321,
  \dodoi{10.1007/BF00649428}

\bibitem[{{Ballantyne}(2004)}]{bal04}
{Ballantyne}, D.~R. 2004, \mnras, 351, 57,
  \dodoi{10.1111/j.1365-2966.2004.07767.x}

\bibitem[{{Ballantyne} {et~al.}(2012){Ballantyne}, {Purvis}, {Strausbaugh}, \&
  {Hickox}}]{bal12}
{Ballantyne}, D.~R., {Purvis}, J.~D., {Strausbaugh}, R.~G., \& {Hickox}, R.~C.
  2012, \apjl, 747, L35, \dodoi{10.1088/2041-8205/747/2/L35}

\bibitem[{{Ballantyne} {et~al.}(2001){Ballantyne}, {Ross}, \& {Fabian}}]{bal01}
{Ballantyne}, D.~R., {Ross}, R.~R., \& {Fabian}, A.~C. 2001, \mnras, 327, 10,
  \dodoi{10.1046/j.1365-8711.2001.04432.x}

\bibitem[{{Ballantyne} {et~al.}(2002){Ballantyne}, {Ross}, \& {Fabian}}]{bal02}
---. 2002, \mnras, 336, 867, \dodoi{10.1046/j.1365-8711.2002.05818.x}

\bibitem[{{Ballantyne} {et~al.}(2004){Ballantyne}, {Turner}, \&
  {Blaes}}]{bal04b}
{Ballantyne}, D.~R., {Turner}, N.~J., \& {Blaes}, O.~M. 2004, \apj, 603, 436,
  \dodoi{10.1086/381495}

\bibitem[{{Ballantyne} {et~al.}(2005){Ballantyne}, {Turner}, \&
  {Young}}]{bal05}
{Ballantyne}, D.~R., {Turner}, N.~J., \& {Young}, A.~J. 2005, \apj, 619, 1028,
  \dodoi{10.1086/426578}

\bibitem[{{Brenneman}(2013)}]{bre13b}
{Brenneman}, L. 2013, {Measuring the Angular Momentum of Supermassive Black
  Holes}, \dodoi{10.1007/978-1-4614-7771-6}

\bibitem[{{Buisson} {et~al.}(2019){Buisson}, {Fabian}, {Barret}, {F{\"u}rst},
  {Gandhi}, {Garc{\'\i}a}, {Kara}, {Madsen}, {Miller}, {Parker}, {Shaw},
  {Tomsick}, \& {Walton}}]{bui19}
{Buisson}, D.~J.~K., {Fabian}, A.~C., {Barret}, D., {et~al.} 2019, \mnras, 490,
  1350, \dodoi{10.1093/mnras/stz2681}

\bibitem[{{Chandrasekhar}(1960)}]{cha60}
{Chandrasekhar}, S. 1960, {Radiative transfer} (New York: Dover)

\bibitem[{{Dirac}(1925)}]{dir25}
{Dirac}, P.~A.~M. 1925, \mnras, 85, 825, \dodoi{10.1093/mnras/85.8.825}

\bibitem[{{Done} \& {Gierli{\'n}ski}(2006)}]{don06}
{Done}, C., \& {Gierli{\'n}ski}, M. 2006, \mnras, 367, 659,
  \dodoi{10.1111/j.1365-2966.2005.09968.x}

\bibitem[{{Fabian} {et~al.}(2017){Fabian}, {Lohfink}, {Belmont}, {Malzac}, \&
  {Coppi}}]{fab17}
{Fabian}, A.~C., {Lohfink}, A., {Belmont}, R., {Malzac}, J., \& {Coppi}, P.
  2017, \mnras, 467, 2566, \dodoi{10.1093/mnras/stx221}

\bibitem[{{Fabian} {et~al.}(2015){Fabian}, {Lohfink}, {Kara}, {Parker},
  {Vasudevan}, \& {Reynolds}}]{fab15}
{Fabian}, A.~C., {Lohfink}, A., {Kara}, E., {et~al.} 2015, \mnras, 451, 4375,
  \dodoi{10.1093/mnras/stv1218}

\bibitem[{{Fabian} {et~al.}(1989){Fabian}, {Rees}, {Stella}, \&
  {White}}]{fab89}
{Fabian}, A.~C., {Rees}, M.~J., {Stella}, L., \& {White}, N.~E. 1989, \mnras,
  238, 729

\bibitem[{{Garc{\'{\i}}a} {et~al.}(2013){Garc{\'{\i}}a}, {Dauser}, {Reynolds},
  {Kallman}, {McClintock}, {Wilms}, \& {Eikmann}}]{gar13a}
{Garc{\'{\i}}a}, J., {Dauser}, T., {Reynolds}, C.~S., {et~al.} 2013, \apj, 768,
  146, \dodoi{10.1088/0004-637X/768/2/146}

\bibitem[{{Garc{\'{\i}}a} \& {Kallman}(2010)}]{gar10}
{Garc{\'{\i}}a}, J., \& {Kallman}, T.~R. 2010, \apj, 718, 695,
  \dodoi{10.1088/0004-637X/718/2/695}

\bibitem[{{Garc{\'{\i}}a} {et~al.}(2011){Garc{\'{\i}}a}, {Kallman}, \&
  {Mushotzky}}]{gar11}
{Garc{\'{\i}}a}, J., {Kallman}, T.~R., \& {Mushotzky}, R.~F. 2011, \apj, 731,
  131, \dodoi{10.1088/0004-637X/731/2/131}

\bibitem[{{Garc{\'{\i}}a} {et~al.}(2014){Garc{\'{\i}}a}, {Dauser}, {Lohfink},
  {Kallman}, {Steiner}, {McClintock}, {Brenneman}, {Wilms}, {Eikmann},
  {Reynolds}, \& {Tombesi}}]{gar14a}
{Garc{\'{\i}}a}, J., {Dauser}, T., {Lohfink}, A., {et~al.} 2014, \apj, 782, 76,
  \dodoi{10.1088/0004-637X/782/2/76}

\bibitem[{{Garc{\'{\i}}a} {et~al.}(2016){Garc{\'{\i}}a}, {Fabian}, {Kallman},
  {Dauser}, {Parker}, {McClintock}, {Steiner}, \& {Wilms}}]{gar16b}
{Garc{\'{\i}}a}, J.~A., {Fabian}, A.~C., {Kallman}, T.~R., {et~al.} 2016,
  \mnras, 462, 751, \dodoi{10.1093/mnras/stw1696}

\bibitem[{{Garc{\'{\i}}a} {et~al.}(2018){Garc{\'{\i}}a}, {Kallman}, {Bautista},
  {Mendoza}, {Deprince}, {Palmeri}, \& {Quinet}}]{gar18a}
{Garc{\'{\i}}a}, J.~A., {Kallman}, T.~R., {Bautista}, M., {et~al.} 2018, in
  Astronomical Society of the Pacific Conference Series, Vol. 515, Workshop on
  Astrophysical Opacities, ed. C.~{Mendoza}, S.~{Turck-Chi\`eze}, \&
  J.~{Colgan} (San Francisco: Astronomical Society of the Pacific), 282--288

\bibitem[{{Garc{\'{\i}}a} {et~al.}(2015){Garc{\'{\i}}a}, {Steiner},
  {McClintock}, {Remillard}, {Grinberg}, \& {Dauser}}]{gar15a}
{Garc{\'{\i}}a}, J.~A., {Steiner}, J.~F., {McClintock}, J.~E., {et~al.} 2015,
  \apj, 813, 84, \dodoi{10.1088/0004-637X/813/2/84}

\bibitem[{{George} \& {Fabian}(1991)}]{geo91}
{George}, I.~M., \& {Fabian}, A.~C. 1991, \mnras, 249, 352

\bibitem[{{Guilbert}(1981)}]{guilbert1981}
{Guilbert}, P.~W. 1981, \mnras, 197, 451, \dodoi{10.1093/mnras/197.2.451}

\bibitem[{{Guilbert} \& {Rees}(1988)}]{gui88}
{Guilbert}, P.~W., \& {Rees}, M.~J. 1988, \mnras, 233, 475

\bibitem[{{Haardt}(1993)}]{haa93}
{Haardt}, F. 1993, \apj, 413, 680, \dodoi{10.1086/173036}

\bibitem[{{Hua} \& {Titarchuk}(1995)}]{hua95}
{Hua}, X.-M., \& {Titarchuk}, L. 1995, \apj, 449, 188, \dodoi{10.1086/176045}

\bibitem[{{Hubeny} \& {Mihalas}(2014)}]{hub14}
{Hubeny}, I., \& {Mihalas}, D. 2014, {Theory of Stellar Atmospheres}
  (Princeton, NJ: Princeton University Press)

\bibitem[{Hunter(2007)}]{hun07}
Hunter, J.~D. 2007, Computing in Science \& Engineering, 9, 90,
  \dodoi{10.1109/MCSE.2007.55}

\bibitem[{{Jones}(1968)}]{jon1968}
{Jones}, F.~C. 1968, Physical Review, 167, 1159,
  \dodoi{10.1103/PhysRev.167.1159}

\bibitem[{{Kallman} \& {Bautista}(2001)}]{kal01}
{Kallman}, T., \& {Bautista}, M. 2001, \apjs, 133, 221, \dodoi{10.1086/319184}

\bibitem[{{Kara} {et~al.}(2017){Kara}, {Garc{\'{\i}}a}, {Lohfink}, {Fabian},
  {Reynolds}, {Tombesi}, \& {Wilkins}}]{kar17}
{Kara}, E., {Garc{\'{\i}}a}, J.~A., {Lohfink}, A., {et~al.} 2017, \mnras, 468,
  3489, \dodoi{10.1093/mnras/stx792}

\bibitem[{{Kershaw} {et~al.}(1986){Kershaw}, {Prasad}, \& {Beason}}]{ker1986}
{Kershaw}, D.~S., {Prasad}, M.~K., \& {Beason}, J.~D. 1986, \jqsrt, 36, 273,
  \dodoi{10.1016/0022-4073(86)90050-6}

\bibitem[{{Klein} \& {Nishina}(1929)}]{kle29}
{Klein}, O., \& {Nishina}, T. 1929, Zeitschrift fur Physik, 52, 853,
  \dodoi{10.1007/BF01366453}

\bibitem[{{Kolehmainen} {et~al.}(2011){Kolehmainen}, {Done}, \& {D{\'\i}az
  Trigo}}]{kol11}
{Kolehmainen}, M., {Done}, C., \& {D{\'\i}az Trigo}, M. 2011, \mnras, 416, 311,
  \dodoi{10.1111/j.1365-2966.2011.19040.x}

\bibitem[{{Laor}(1991)}]{lao91}
{Laor}, A. 1991, \apj, 376, 90, \dodoi{10.1086/170257}

\bibitem[{{Lightman} {et~al.}(1981){Lightman}, {Lamb}, \& {Rybicki}}]{lig81}
{Lightman}, A.~P., {Lamb}, D.~Q., \& {Rybicki}, G.~B. 1981, \apj, 248, 738,
  \dodoi{10.1086/159198}

\bibitem[{{Lightman} \& {White}(1988)}]{lig88}
{Lightman}, A.~P., \& {White}, T.~R. 1988, \apj, 335, 57,
  \dodoi{10.1086/166905}

\bibitem[{{Madej} \& {R{\'o}za{\'n}ska}(2000)}]{madej2000}
{Madej}, J., \& {R{\'o}za{\'n}ska}, A. 2000, \aap, 363, 1055

\bibitem[{{Madej} \& {R{\'o}{\.z}a{\'n}ska}(2004)}]{madej2004}
{Madej}, J., \& {R{\'o}{\.z}a{\'n}ska}, A. 2004, \mnras, 347, 1266,
  \dodoi{10.1111/j.1365-2966.2004.07310.x}

\bibitem[{{Madej} {et~al.}(2017){Madej}, {R{\'o}{\.z}a{\'n}ska}, {Majczyna}, \&
  {Nale{\.z}yty}}]{mad17}
{Madej}, J., {R{\'o}{\.z}a{\'n}ska}, A., {Majczyna}, A., \& {Nale{\.z}yty}, M.
  2017, \mnras, 469, 2032, \dodoi{10.1093/mnras/stx994}

\bibitem[{{Madej} {et~al.}(2019){Madej}, {R\'o\.za\'nska}, {Majczyna}, \&
  {Nale{\.Z}yty}}]{mad19}
{Madej}, J., {R\'o\.za\'nska}, A., {Majczyna}, A., \& {Nale{\.Z}yty}, M. 2019,
  \mnras, 484, 2831, \dodoi{10.1093/mnras/stz167}

\bibitem[{{Magdziarz} \& {Zdziarski}(1995)}]{mag95}
{Magdziarz}, P., \& {Zdziarski}, A.~A. 1995, \mnras, 273, 837

\bibitem[{{Markoff} {et~al.}(2005){Markoff}, {Nowak}, \& {Wilms}}]{mar05}
{Markoff}, S., {Nowak}, M.~A., \& {Wilms}, J. 2005, 635, 1203

\bibitem[{{Matt} {et~al.}(1993){Matt}, {Fabian}, \& {Ross}}]{mat93}
{Matt}, G., {Fabian}, A.~C., \& {Ross}, R.~R. 1993, \mnras, 262, 179

\bibitem[{{Matt} {et~al.}(1991){Matt}, {Perola}, \& {Piro}}]{mat91}
{Matt}, G., {Perola}, G.~C., \& {Piro}, L. 1991, \aap, 247, 25

\bibitem[{{Matt} {et~al.}(1992){Matt}, {Perola}, {Piro}, \& {Stella}}]{mat92}
{Matt}, G., {Perola}, G.~C., {Piro}, L., \& {Stella}, L. 1992, \aap, 257, 63

\bibitem[{{McClintock} {et~al.}(2013){McClintock}, {Narayan}, \&
  {Steiner}}]{mcc13}
{McClintock}, J.~E., {Narayan}, R., \& {Steiner}, J.~F. 2013, \ssr,
  \dodoi{10.1007/s11214-013-0003-9}

\bibitem[{{McClintock} \& {Remillard}(2006)}]{mcc06}
{McClintock}, J.~E., \& {Remillard}, R.~A. 2006, {Black hole binaries}
  (Cambridge University Press, London), 157--213

\bibitem[{{Mihalas}(1978)}]{mih78}
{Mihalas}, D. 1978, {Stellar atmospheres} (2nd ed.; San Francisco, CA: Freeman)

\bibitem[{{M{\"u}nch}(1948)}]{mun48}
{M{\"u}nch}, G. 1948, \apj, 108, 116, \dodoi{10.1086/145048}

\bibitem[{{Nagirner} \& {Poutanen}(1994)}]{napo1994}
{Nagirner}, D.~I., \& {Poutanen}, J. 1994, {Single Compton scattering}, Vol.~9
  (Amsterdam: Harwood Academic Publishers)

\bibitem[{{Nagirner} \& {Poutanen}(1993)}]{nag93}
{Nagirner}, D.~I., \& {Poutanen}, Y.~J. 1993, Astronomy Letters, 19, 262

\bibitem[{{Nayakshin} \& {Kallman}(2001)}]{nay01}
{Nayakshin}, S., \& {Kallman}, T.~R. 2001, \apj, 546, 406,
  \dodoi{10.1086/318250}

\bibitem[{{Nayakshin} {et~al.}(2000){Nayakshin}, {Kazanas}, \&
  {Kallman}}]{nay00}
{Nayakshin}, S., {Kazanas}, D., \& {Kallman}, T.~R. 2000, \apj, 537, 833,
  \dodoi{10.1086/309054}

\bibitem[{{Nied{\'z}wiecki} {et~al.}(2019){Nied{\'z}wiecki}, {Szanecki}, \&
  {Zdziarski}}]{nie19}
{Nied{\'z}wiecki}, A., {Szanecki}, M., \& {Zdziarski}, A.~A. 2019, \mnras, 485,
  2942, \dodoi{10.1093/mnras/stz487}

\bibitem[{Oliphant(2006)}]{oli06}
Oliphant, T.~E. 2006, A guide to NumPy, Vol.~1 (Trelgol Publishing USA)

\bibitem[{{Poutanen} {et~al.}(1996){Poutanen}, {Nagendra}, \&
  {Svensson}}]{pou96}
{Poutanen}, J., {Nagendra}, K.~N., \& {Svensson}, R. 1996, \mnras, 283, 892

\bibitem[{{Remillard} \& {McClintock}(2006)}]{rem06}
{Remillard}, R.~A., \& {McClintock}, J.~E. 2006, \araa, 44, 49,
  \dodoi{10.1146/annurev.astro.44.051905.092532}

\bibitem[{{Reynolds}(2013)}]{rey13}
{Reynolds}, C.~S. 2013, Classical and Quantum Gravity, 30, 244004,
  \dodoi{10.1088/0264-9381/30/24/244004}

\bibitem[{{Reynolds}(2019)}]{rey19}
---. 2019, Nature Astronomy, 3, 41, \dodoi{10.1038/s41550-018-0665-z}

\bibitem[{{Ross}(1978)}]{ros78PhDT}
{Ross}, R.~R. 1978, PhD thesis, Colorado Univ., Boulder.

\bibitem[{{Ross} \& {Fabian}(1993)}]{ros93}
{Ross}, R.~R., \& {Fabian}, A.~C. 1993, \mnras, 261, 74

\bibitem[{{Ross} \& {Fabian}(2005)}]{ros05}
---. 2005, \mnras, 358, 211, \dodoi{10.1111/j.1365-2966.2005.08797.x}

\bibitem[{{Ross} \& {Fabian}(2007)}]{ros07}
---. 2007, \mnras, 381, 1697, \dodoi{10.1111/j.1365-2966.2007.12339.x}

\bibitem[{{Ross} {et~al.}(1978){Ross}, {Weaver}, \& {McCray}}]{ros78}
{Ross}, R.~R., {Weaver}, R., \& {McCray}, R. 1978, \apj, 219, 292,
  \dodoi{10.1086/155776}

\bibitem[{{R{\'o}{\.z}a{\'n}ska} \& {Madej}(2008)}]{rozanska2008}
{R{\'o}{\.z}a{\'n}ska}, A., \& {Madej}, J. 2008, \mnras, 386, 1872,
  \dodoi{10.1111/j.1365-2966.2008.13173.x}

\bibitem[{{R{\'o}{\.z}a{\'n}ska} {et~al.}(2011){R{\'o}{\.z}a{\'n}ska}, {Madej},
  {Konorski}, \& {SaḐowski}}]{rozanska2011}
{R{\'o}{\.z}a{\'n}ska}, A., {Madej}, J., {Konorski}, P., \& {SaḐowski}, A.
  2011, \aap, 527, A47, \dodoi{10.1051/0004-6361/201015626}

\bibitem[{{Shakura} \& {Sunyaev}(1973)}]{sha73}
{Shakura}, N.~I., \& {Sunyaev}, R.~A. 1973, \aa, 24, 337

\bibitem[{{Sunyaev} \& {Titarchuk}(1980)}]{sutit1980}
{Sunyaev}, R.~A., \& {Titarchuk}, L.~G. 1980, \aap, 500, 167

\bibitem[{Thomson(1906)}]{thom1906}
Thomson, J. 1906, Conduction of Electricity Through Gases, Cambridge physical
  series (University Press).
\newblock \url{https://books.google.de/books?id=k1ZKAAAAMAAJ}

\end{thebibliography}
%
%==================================================================================
%
%
%
\end{document}